\documentclass[preprint,review,12pt]{elsarticle}
\usepackage{latexsym}
\usepackage{natbib}
\usepackage{subfig}
\usepackage[percent]{overpic} %esto es para para enumerar a las subfiguras: (a),(b),etc.
\usepackage{float}  %esto es para ubicar las figuras donde yo quiero. Hay que poner [H] y no [h] como hacía siempre.
\usepackage{appendix} %sirve para poner apéndice
\usepackage{booktabs} %booktabs es para hacer tablas pro.
\usepackage{natbib}
\usepackage{multirow}%es para poder unir filas en una tabla
\usepackage{array}
\usepackage{amsmath}%permite usar las abreviaciones de los simbolos como dfrac
\usepackage{amsfonts}%permite escribir con los simbolos matematicos, las fuentes de la american mathematical society
\usepackage{amssymb}%permite declarar nuevos comandos para simbolos: \DeclareMathSymbol
\usepackage[T1]{fontenc}
\usepackage{verbatim}   % useful for program listings
\usepackage{t1enc}
\usepackage{anysize}
\usepackage{tocbibind}

\newcommand{\etal}{\textit{et al.}}
\newcommand{\ie}{\textit{i.e.}}
\begin{document}

\begin{frontmatter}

\title{Temporal percolation of a susceptible adaptive network} 

\author[mdp]{L. D. Valdez\corref{cor}}
\ead{ldvaldes.at.mdp.edu.ar}
\author[mdp]{P. A. Macri}
\author[mdp,bst]{L. A. Braunstein}
\cortext[cor]{Corresponding author}

\address[mdp]{Instituto de Investigaciones F\'isicas de Mar del Plata
  (IFIMAR)-Departamento de F\'isica, Facultad de Ciencias Exactas y
  Naturales, Universidad Nacional de Mar del Plata-CONICET, Funes
  3350, (7600) Mar del Plata, Argentina.}
\address[bst]{Center for Polymer Studies, Physics Department, Boston
  University, Boston, Massachusetts 02215, USA.}

\begin{abstract}
In the last decades, many authors have used the
susceptible-infected-recovered model to study the impact of the
disease spreading on the evolution of the infected
individuals. However, few authors focused on the temporal unfolding of
the susceptible individuals. In this paper, we study the dynamic of
the susceptible-infected-recovered model in an adaptive network that
mimics the transitory deactivation of permanent social contacts, such
as friendship and work-ship ties. Using an edge-based compartmental
model and percolation theory, we obtain the evolution equations for
the fraction susceptible individuals in the susceptible biggest
component. In particular, we focus on how the individual's behavior
impacts on the dilution of the susceptible network. We show that, as a
consequence, the spreading of the disease slows down, protecting the
biggest susceptible cluster by increasing the critical time at which
the giant susceptible component is destroyed. Our theoretical results
are fully supported by extensive simulations.
\end{abstract}
\begin{keyword}
Epidemic Models \sep   Percolation \sep   Adaptive networks
\end{keyword}
\end{frontmatter}

\section{Introduction}
Dynamic topologies in complex network models have recently become a
subject of intensive investigations~\cite{Gro_01,Gro_04}. While in the
past, many dynamical processes were mainly developed numerically and
analytically on static networks, most of the real networks alter their
topologies over time. As a consequence, recently many researchers are
modeling these processes on top of dynamic networks. The topology of
the networks can change by evolution or by adaptive
dynamic~\cite{Gro_01,Gro_04}. Those networks in which the structure
changes regardless of the processes taking place on top of them are
called evolutive networks. On the other hand, networks that alter
their topology to mitigate or promote these processes are called
adaptive networks~\cite{Gro_01,Sch_02,Vaz_02}. In adaptive networks,
there is a co-evolution between the dynamic process and the
topology. The study of adaptive networks has generated a great
interest in many disciplines since there is evidence of many real
systems with an adaptive topology~\cite{Gro_01}. For example, from the
analysis of the gene regulatory networks, it was shown that
interactions between genes can change in response to diverse stimuli,
leading to changes in the network's topology~\cite{Lus_01,Gro_01}. On
the other hand, the recent pandemics of SARS~\cite{Col_01} and
H1N1~\cite{Baj_01}, have promoted the modeling of adaptive strategies
on the contact network to slow down the spread of the
epidemics~\cite{Gro_02,Fun_01,Fen_01,Mao_01,Wan_01,Tun_01}. The study
of these mitigation strategies could provide important information to
adopt health policies, allowing to characterize the effect of the
strategy on the structure of the society and to quantify its
effectiveness against the spread of an epidemic.

The most used model that represents these recent diseases is the
susceptible-infected-recovered (SIR) model~\cite{Boc_01,And_01}, in
which the individuals can be in one of three states, susceptible,
infected or recovered. In this model, the disease propagates on top of
the contact network until it reaches the steady state, \ie , when
there are no more infected individuals. In static networks, the steady
state of the SIR model was widely studied using percolation tools
through a generating function
formalism~\cite{Gra_01,New_05,Mil_01,Mey_01}. It is known that the
final size of the disease (fraction of recovered individuals) is
governed by a control parameter which is the effective probability of
infection or transmissibility $T$ of the disease. At a critical
threshold $T=T_c$, the disease overcomes a second order phase
transition with an epidemic phase for $T>T_c$, while for $T<T_c$ the
disease consists only of outbreaks that reach a small fraction of the
population. 

In the last decade, there has been considerable progress to described
theoretically the dynamics of the disease spreading of this model in
static networks~\cite{Vol_01,Mil_03,Mil_04,Lin_01,KarNew_01}. However
these researches have only focused on the description of the temporal
evolution of compartmental quantities, such us the fraction of the
infected and susceptible individuals. Recently, Valdez
\etal~\cite{Val_02} using an edge-based compartmental model
(EBCM)~\cite{Vol_01,Mil_03,Mil_04} and a generating function
formalism, proposed a set of equations which describe the evolution of
the size of the giant component of the susceptible individuals (GSC)
in static networks. They found that a time-dependent quantity
$\Phi_S(t)$, namely, the probability that a given neighbor of a root
node is susceptible at time $t$ (more details are given below), is the
control parameter of a temporal node void percolation process
involving the network composed by susceptible individuals.  They
showed that the GSC overcomes a second order phase transition at a
critical time $t_c$ above which it is destroyed.

In contrast to the SIR model in static networks, the study of this
model on adaptive topologies has being explored by few authors. Wang
\etal~\cite{Wan_01} proposed an intervention strategy in which
susceptible individuals, induced by fear, break the links with their
neighbors, with a probability related to the number of infected
neighbors, regardless of the state of the neighbors. They show that
this strategy decreases the number of infected individuals and delays
the progression of the disease compared to the case of
non-intervention. Lagorio \etal~\cite{Lag_01} studied a rewiring
strategy in which susceptible individuals redirect their links with
infected neighbors towards other susceptible individuals with
probability $w$. They showed that there is a phase transition at a
critical rewiring threshold $w_{c}$ separating an epidemic from a
non-epidemic phase, which can be related to static link
percolation. In a recent paper~\cite{Val_01}, it was proposed an
adaptive SIR model driven by intermittent connections where the
susceptible individuals, using local information, break the links with
their infected neighbors with probability $\sigma$ for an interval
$t_b$ after which they reestablish the connections with their previous
contacts. This model, called intermittent social distancing (ISD)
strategy focus to model the behavior of individuals who preserve their
closer contacts during the disease spreading. Using the framework of
percolation theory, they derived the transmissibility $T_{\sigma}$ and
found that there exists a critical probability $\sigma_{c}$ depending
on $t_b$, above which the epidemic spread is stopped. They also showed
that the ISD strategy, produces a ``susceptible herd behavior'' below
a transmissibility $T^{*}$~\cite{Val_01,New_06} that protects a large
cluster of susceptible individuals from being infected. This focus on
the susceptible network provides a description of the functional
network, since the GSC is the one that supports the economy of a
society.

Until now, these researches have focused on the effect of the
strategies on quantities in the steady state and very little has made
to describe the dynamics. Moreover, so far less explored, is the study
of the effect of these strategies on the GSC.

In this paper, using the EBCM approach and percolation theory, we
study the temporal evolution of the fraction of the susceptible
individuals on adaptive networks following the ISD strategy. We find
that this strategy protects a giant susceptible cluster by increasing
the time $t_c$ at which the functional network is destroyed. In a more
realistic scenario where the implementation of the strategy is
delayed, we obtain that it also increases $t_c$, and find that the
dilution of the GSC can also be described theoretically by percolation
tools. The paper is organized as following: in Sec.~\ref{Node_Void},
we explain the node void percolation that describes the dilution of
the susceptible network. In Sec.~\ref{Sec_Inf_Cur} we derive the
dynamic equations of the ISD strategy and in Sec.~\ref{Sec_EvolSusCl}
we present the theoretical and the simulation results of the ISD
strategy implemented from the beginning of the disease spreading. In
Sec.~\ref{SecDel_Straq} we derive the evolution equations when the
implementation of the strategy is delayed. Finally, in
Sec.~\ref{SecConc} we present our conclusions and outlooks.

\section{Framework of Node Void Percolation}\label{Node_Void}
In the discrete version of the SIR model, an infected individual
infects a susceptible neighbor with probability $\beta$ and he
recovers after $t_r$ time units since he was infected, where $t_r$ is
called the recovery time. It is known that in this model the control
parameter is the transmissibility $T=1-(1-\beta)^{t_r}$ that governs
the final fraction of infected individuals. This model can be mapped
into link percolation since when the disease traverses a link with a
transmissibility $T$, this process is equivalent to occupy that link
with the same probability $T$.

On the other hand, the network composed by susceptible individuals is
diluted during the epidemic spreading, since when the disease
traverses a link, the susceptible individual is removed from the
susceptible network. However, this process is different than an
ordinary node percolation because the susceptible nodes are not chosen
randomly. Instead, they are reached by the disease following a link,
and therefore, nodes with higher connectivity are reached with higher
probability than nodes with lower connectivity. This kind of dilution,
that we called node void percolation~\cite{Val_02} leads, in the
steady state, to a second order phase transition in which the order
parameter is the fraction of nodes $S_{1}$ in the GSC, and the control
parameter is $\Phi_{S}(\infty)$, \ie, the probability that following a
random chosen link, a susceptible node is reached. Moreover, since the
susceptible network loses the nodes with the highest connectivities,
the phase transition at $\Phi_{S}(\infty)=\Phi_{Sc}$, has mean field
exponents, as in an intentional attack process independently of the
network's topology~\cite{Coh_04,Val_02}.

When $T$ increases, $S_{1}$ decreases and therefore $\Phi_{S}(\infty)$ and
$T$ are interrelated. In particular, there is an effective
transmissibility $T^{*}$ that set the critical value of $\Phi_{Sc}$ at
which $S_{1}\to 0$, and its value can be obtained using a generating
function formalism~\cite{Val_01,New_06}. Denoting the degree
distribution as $P(k)$, $T^{*}$ fulfills the equation
\begin{equation}\label{TestrellaFinal}
\Phi_{Sc}=G_{1}[1-T^{*}(1-\Phi_{Sc})],
\end{equation}
where $T^{*}$ is the solution of Eq.~(\ref{TestrellaFinal}),
$\Phi_{Sc}=G_{1}\left[(G_{1}^{'})^{-1}(1)\right]$ and
$G_{1}(x)=\sum_{k}kP(k)/\langle k \rangle x^{k-1}$ is the generating
function for the excess degree distribution, that is, the degree
distribution of the remaining outgoing links in a branching
process~\cite{New_07, Exxon_06}.

\section{Dynamic framework of the ISD strategy}\label{Sec_Inf_Cur}
In the ISD model, initially all the nodes are susceptible except for
one node randomly infected, that represents the patient zero from
which the disease spreads. In this paper, we call active links to the
links between infected and susceptible individuals. An infected
individual transmits the disease to a susceptible neighbor with
probability $\beta$ and, if he fails, the susceptible individual
breaks his link with the infected one with probability $\sigma$ for
$t_b$ time units. After a period $t_b$ both individuals are
reconnected and the process is repeated until the infected individual
recovers at a fixed time $t_r>t_b$ since he was infected. If an active
link is broken for more than $t_b=t_r-1$ time units, it is restored as
a non-active link since the infected individual is recovered. In this
model, the disease spreads with an effective transmissibility
$T(\beta,\sigma,t_r,t_b)\equiv T_{\sigma}$~\cite{Val_01} given by,
\begin{eqnarray}\label{Ec.Trans}
T_{\sigma}=\sum_{m=1}^{t_r} \beta(1-\beta)^{m-1}(1-\sigma)^{m-1}+ \beta
\sum_{m=t_{b}+2}^{t_r}\phi(m,t_b,\sigma,\beta).
\end{eqnarray}
In the first term of Eq.~(\ref{Ec.Trans}),
$\beta(1-\beta)^{m-1}(1-\sigma)^{m-1}$ is the probability that an
active link is lost due to the infection of the susceptible individual
at time $m$ given that the active link has never been broken in the
$m-1$ steps since it appears. In the second term of
Eq.~(\ref{Ec.Trans}), $\beta\;\phi(m,t_b,\sigma,\beta)$ denotes the
probability that an active link is lost due to the infection of the
susceptible individual at time $m$ given that the link was broken at
least once in the first $m-1$ time units. The probability
$\phi(m,t_b,\sigma,\beta)$, which is only valid for $m\geq t_b+2$ is
given by~\cite{Val_01}
\begin{equation}\label{eqSumPhi}
\phi(m,t_b,\sigma,\beta)\equiv\phi_{m} =
\sum_{u=1}^{\left[\frac{m-1}{t_{b}+1}\right]}\binom{m-u\;t_{b}-1}{u}\sigma^{u}(1-\sigma)^{m-1-u(t_{b}+1)}(1-\beta)^{m-1-u\;t_{b}},
\end{equation}
where $\left[\cdots \right]$ denotes the integer part
function. Thus, the ISD strategy~\cite{Val_01} reduces the initial
transmissibility $T$ towards an effective transmissibility
$T_{\sigma}$, mitigating the spread of the disease.

Eq.~(\ref{Ec.Trans}) can be rewritten as
\begin{eqnarray}\label{Ec.Trans_sigma}
T_{\sigma}=\beta\sum_{m=1}^{t_r}\Omega_{m},
\end{eqnarray}
with
\begin{eqnarray}\label{Ec.omesuper}
\Omega_{m}\equiv (1-\beta)^{m-1}(1-\sigma)^{m-1}+\phi_m,
\end{eqnarray}
where $\Omega_{m}$ is the probability that an active link has neither
ever transmitted the disease nor been broken during the first $m-1$
time units, plus the probability that the active link was broken at
least once in the first $m-1$ time units but at time $m$ it is
present. Here, $\phi_m=0$ for $m<t_b+2$ and is given by
Eq.~(\ref{eqSumPhi}) for $t_b+2\leq m\leq t_r$.

In order to study the evolution of the number of infected and
susceptible individuals in the ISD model, we use the edge-based
compartmental approach~\cite{Vol_01,Mil_03,Mil_04} based on a
generating function formalism. Denoting the fraction of susceptible,
infected and recovered individuals at time $t$ by $S(t)$, $I(t)$ and
$R(t)$, respectively, the EBCM approach lies on describing the
evolution of the probability that a randomly chosen node is
susceptible. In order to compute $S(t)$, a link is randomly chosen and
then a direction is given, in which the node in the target of the
arrow is called the root, and the base is its neighbor. Then the
fraction of susceptible individuals is given by $S(t)=\sum_{k}
P(k)\theta_t^{k}=G_{0}(\theta_t)$, where $G_{0}(x)$ is the generating
function of the degree distribution and $\theta(t)\equiv \theta_t$ is
the probability that the base does not transmit the disease to the
root. In this approach, infection from the root to the base is
disallowed, in order to treat the state of the root's neighbors as
independent~\cite{Mil_03,Mil_04}. In order to find $\theta_t$ we have
to compute the probabilities $\Phi_{S}(t)$, $\Phi_{R}(t)$ and
$\Phi_{I}(t)$ that the base is susceptible, recovered or infected but
has not yet transmitted the disease to the root. These probabilities
satisfy the relation
\begin{eqnarray}\label{thetaEqPhi}
\Phi_{S}(t)+\Phi_{I}(t)+\Phi_{R}(t)=\theta_t.
\end{eqnarray}
As $\Phi_S(t)$\footnote{Note that $\Phi_{S}(t)$ can be computed as the
  square root of the fraction of links between susceptible
  individuals, because $[\Phi_{S}(t)]^2$ is the probability that both
  stubs of a random chosen link belong to susceptible nodes.} depends
on the second neighbors of the root, we have to disallow infection
from the first neighbor (base of the root) to the second neighbors,
thus
\begin{eqnarray}\label{defphine}
\Phi_{S}(t)=G_1(\theta_t).
\end{eqnarray}

Then, using the EBCM approach adapted to discrete times, the evolution
of $\Phi_{S}(t)$ and $\Phi_{I}(t)$ in the ISD model is given by
deterministic equations, that are only valid above the critical
transmissibility $T_{c}$ where there is a macroscopic fraction of
infected nodes,
\begin{eqnarray}   
\Delta{\theta_t}&=&-\beta\Phi_{I}(t),\label{delta_theta}\\
\Delta{\Phi}_{S}(t)&=&G_{1}(\theta_{t+1})-G_{1}(\theta_{t}),\label{deltaPhiS}
\end{eqnarray}
where $\Delta$ is the discrete change of the variables between times
$t$ and $t+1$. Eq.~(\ref{delta_theta}) represents the decrease of
$\theta_t$ when the disease traverses an active
link. Eq.~(\ref{deltaPhiS}) describes the decrease of $\Phi_{S}(t)$
(see Eq.~(\ref{defphine})). It is important to note that
$\Delta\Phi_{S}(t)$ is non-positive, and $-\Delta\Phi_{S}(t)$
represents the new active links.

The evolution equation of $\Phi_{I}(t)$ is given by,
\begin{eqnarray}
\Delta{\Phi}_{I}(t)&=&-\beta\Phi_{I}(t)-\Delta{\Phi}_{S}+(1-\beta)(1-\sigma)\;\Omega_{t_{r}}\Delta{\Phi}_{S}(t-t_{r})+\Psi(t),\label{deltaPhi}
\end{eqnarray}
where
\begin{eqnarray}\label{reconection_term}
\Psi(t)&\equiv&-\sigma(1-\beta)\Phi_{I}(t)+\sigma(1-\beta)\Phi_{I}(t-t_{b})\nonumber\\
&&+\sum_{i=0}^{t_{b}}\sigma(1-\beta)\Omega_{t_r-t_b+i}\Delta{\Phi}_{S}(t-i-t_{r}),
\end{eqnarray}
with,
\begin{eqnarray}\label{Eq_Omega}
\Omega_{t_r-t_b+i}=(1-\beta)^{t_{r}-t_b+i-1}(1-\sigma)^{t_{r}-t_b+i-1}+\phi_{t_r-t_b+i},
\end{eqnarray}
already defined by Eq.~(\ref{Ec.omesuper}).

In Eq.~(\ref{deltaPhi}) the first term represents the decrease of
$\Phi_{I}(t)$ when the infected base node transmits the disease to the
root; the second term corresponds to an increasing of $\Phi_{I}(t)$
due to the new infections. The third term corresponds to the recovery
of the infected nodes in active links, which is proportional to: a)
the active links that appeared $t_r$ time units ago \ie,
$\Delta{\Phi}_{S}(t-t_{r})$ and b) to the probability that these
active links are present $t_r-1$ time units after they appeared and in
the last time unit they do not transmit the disease neither being
broken, \ie, $(1-\beta)(1-\sigma)\Omega_{t_{r}}$. Note that
$\Omega_{t_{r}}\Delta{\Phi}_{S}(t-t_{r})$ is related with the active
links that appeared $t_r$ time units ago, and remain active in their
last time unit before the nodes recover and the active links
disappear. The last term corresponds to breaking and reconnection of
active links [see Eq.~(\ref{reconection_term})].

The first term of Eq.~(\ref{reconection_term}), which we call
``breaking term'', represents the decrease of $\Phi_{I}(t)$ due to
breaking active links during a period $t_b$, which is proportional to
$\sigma(1-\beta)$. The remaining terms, which we call ``reconnection
terms'', represent the increase of $\Phi_{I}(t)$ due to reconnecting
active links. While the reconnection term could be thought as minus
the ``breaking term'' delayed by $t_{b}$ time units [\ie,
$\sigma(1-\beta)\Phi(t-t_{b})$], that term needs to be corrected due
to the recovery of some infected individuals that reduce the number of
active links that could be restored at time $t+1$. In
Table~\ref{History} we show a schematic with an example of the correction
term. Then $\Psi(t)$ [see Eq.~(\ref{reconection_term})] represents the
net flow between the decrease of the probability $\Phi_{I}(t)$ due to
breaking active links, and the increase of $\Phi_{I}(t)$ due to
restoring active links that have being disconnected $t_b$ units time
ago.

%{\multicolumn{1}{m{11cm}}
\begin{table}[H]
\caption{Schematic cases of the ``reconnection terms'' of
  Eq.~(\ref{reconection_term}) for $t_{r}=10$ and $t_{b}=4$. The first
  column represents a typical configuration of different cases of the
  ``reconnection term''. The second column is the probability that an
  active link cannot be restored at time $t$. In the first column,
  each cell corresponds to a unit time.  The white cells represent the
  time unit at which an active link between the susceptible and the
  infected node exists, the gray ones represent to the disconnection
  period and the crosses correspond to the case where the infected is
  recovered. The moment at which an active link appears is represented by
  (I), and the moment when this pair, at time $t-t_b$, breaks for
  the next $t_b$ time units is represented by (II). When some active
  links break at time $t-t_{b}$ at the beginning of time $t$ they
  have to be restored, which is the case of 1). However, other active
  links which break at time $t-t_{b}$ cannot be restored as active
  links because the infected node recovers. This is shown in the cases
  2a) and 2b) where the infected node has recovered at $i=2$ time
  units before $t$ (where $i$ is the sub-index of
  Eq.~(\ref{Eq_Omega})). The case 2a), corresponds to an active link
  which appears at time $t-t_r-i=t-12$ and breaks by the first
  time $8$ times unit later. The probability for this case is
  $\sigma(1-\beta)^{8}(1-\sigma)^{7}$. Similarly, the case 2b)
  corresponds to an active link which appears at time
  $t-t_r-i=t-12$ and breaks, but not by the first time, $8$ time
  units later. The probability of this case is
  $\sigma(1-\beta)\phi_{8}$.}  \centering
\renewcommand{\arraystretch}{1}
%esto es para aumentar la separaci\'on entre filas
\begin{tabular}{c@{\hspace{1cm}}c@{\hspace{0.5cm}}}  %% el @{} es para agregar espacio
\toprule[0.05cm]
\multicolumn{1}{c}{Example}&\multicolumn{1}{m{6cm}}{Probability that the active link cannot be restored at time $t$}\\
\cmidrule [0.03cm]{1-2}
\multicolumn{1}{m{6cm}} {\includegraphics[scale=0.50]{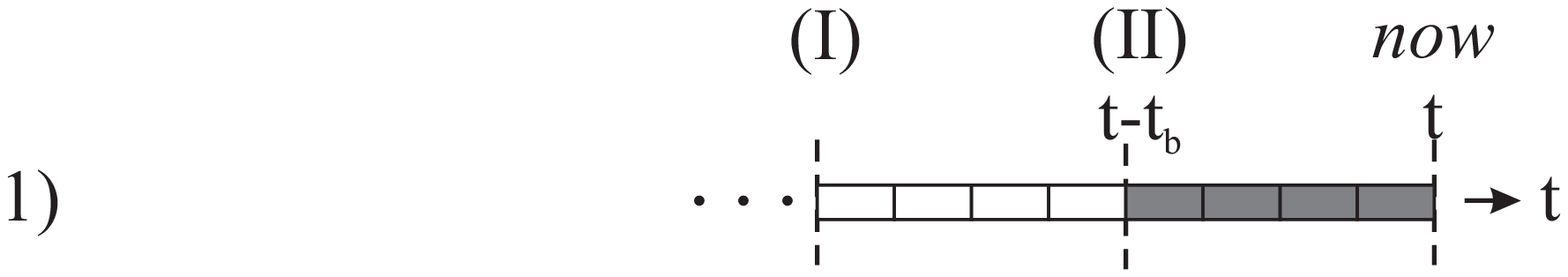}}&\multicolumn{1}{c}{0}\\\cmidrule{1-2}
\multirow{6}{9cm}{\includegraphics[scale=0.50]{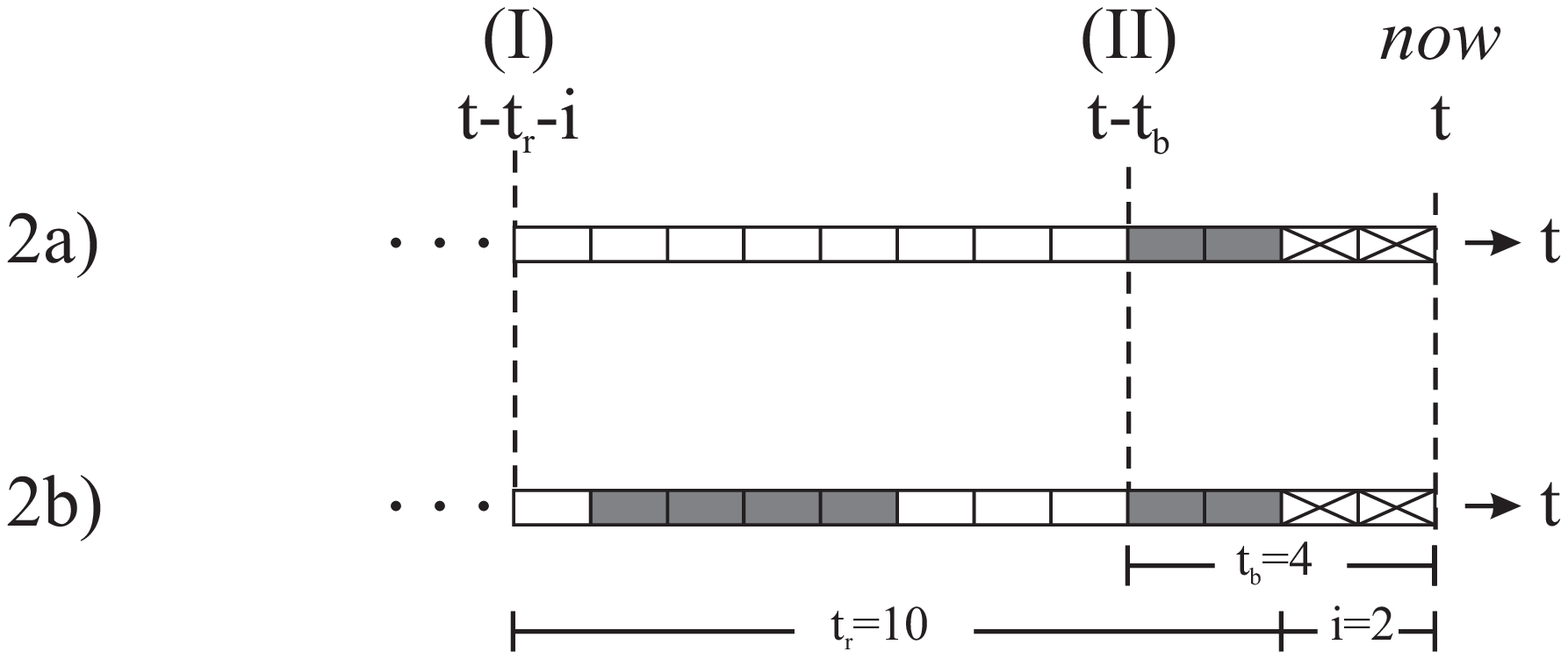}}&\\\addlinespace\addlinespace&\multicolumn{1}{c}{$\sigma(1-\beta)^{8}(1-\sigma)^{7}$}\\&\\&\multicolumn{1}{c}{$\sigma(1-\beta)\phi_{8}$}\\&\\&\\
\bottomrule[0.05cm]
\end{tabular}
\label{History}
\end{table}

Note that in Eq.~(\ref{reconection_term}), $(1-\beta)\sigma$ is the
probability that an active link breaks. By contrast, $\Phi_{I}(t)$
corresponds to the state of the base node and no information is
provided about the state of the root node. Therefore,
Eqs.~(\ref{deltaPhi})~and~(\ref{reconection_term}) correspond to a
process wherein the infected nodes break their links with infected and
recovered individuals too. However, the breaking of these links has no
effect on the evolution of $I(t)$ and $S(t)$ because those links are not
able to transmit the disease. Therefore,
Eqs.~(\ref{deltaPhi})~and~(\ref{reconection_term}) are also valid for
our ISD strategy, where only the susceptible individuals break
their links with the infected ones.

The evolution of the fraction of infected individuals is given
by~\cite{Val_02}
\begin{eqnarray}\label{EqInfec}
\Delta{I}(t)&=& -\Delta S(t)+\Delta S(t-t_r),
\end{eqnarray}
where $-\Delta S(t)=-(G_{0}(\theta_{t+1})-G_{0}(\theta_{t}))$, as
explained above, represents the fraction of new infected
individuals. The second term represents the decrease of $I(t)$ due to
the recovery of infected individuals that have been infected $t_{r}$
time units ago. Note that $I(t)$ contains the new infected individuals
at time $t$, and also the infected in a previous time.

The evolution of the fraction of susceptible individuals in the
susceptible giant component $S_{1}(t)$ is obtained~\cite{Val_02} by
solving the equations
\begin{eqnarray}
S_{1}(t)&=&G_{0}(\theta_{t})-G_{0}(\omega_{t}),\label{EqSusTotal2}\\
\omega_{t}&=&\theta_{t}-G_{1}(\theta_{t})+G_{1}(\omega_{t}),\label{EqSusTotal1}
\end{eqnarray}
where $\omega_t$ is the probability that a base individual, which is
not connected to the giant susceptible cluster, has not yet
transmitted the disease to the root at time $t$;
$G_{0}(\theta_t)=S(t)$ is the total fraction of susceptible
individuals and $G_{0}(\omega_t)$ is the fraction of individuals
belonging to finite susceptible clusters. In this paper, we only
present our theoretical and numerical results for the susceptible
nodes, because we are interested here only on the functional
network. However, we checked that Eq.~(\ref{EqInfec}) is in fully
agreement with the simulations for the ISD strategy.

\section{Theoretical and simulation results}\label{Sec_EvolSusCl}

We apply the ISD strategy on Erd\"os R\'enyi networks (ER) with degree
distribution $P(k)=\langle k \rangle^{k}\exp(-\langle k \rangle)/k!$,
in which $k$ is the connectivity and $\langle k \rangle$ is the average
degree of the network, and Scale-Free networks (SF) with degree
distribution $P(k)\sim k^{-\lambda}$ where $k_{min}\leq k \leq
k_{max}$ and $\lambda$ is a measure of the heterogeneity of the degree
distribution.

In order to obtain theoretically $S_{1}(t)$, we iterate Eqs.
(\ref{delta_theta})-(\ref{EqSusTotal2}) with initial condition
$\theta_{0}=1-1/N$, $\Phi_{I}(0)=1/N$ where $N$ is the number of nodes
in the network. Since at the beginning of the spreading the fraction
of infected individuals is very small, the time $t$ is shifted to
$t=0$ when $1$\% of the individual are infected~\cite{Vol_01} because
this stochastic regime is not taken into account in our deterministic
equations.

For the numerical simulations we construct the networks using the
configurational model ~\cite{Mol_01}, and for the disease propagation
we use the algorithm explained in Sec.~\ref{Sec_Inf_Cur}. For SF
networks, we choose $k_{min}=2$ in order to ensure that the network is
fully connected~\cite{Han_01}.  In the simulations, the disease has a
transmissibility $T$ without strategy, but it will spread with
transmissibility $T_{\sigma}$ [see Eq.~(\ref{Ec.Trans})] due to the
ISD strategy that is applied since the patient zero appears.

\begin{figure}[H]
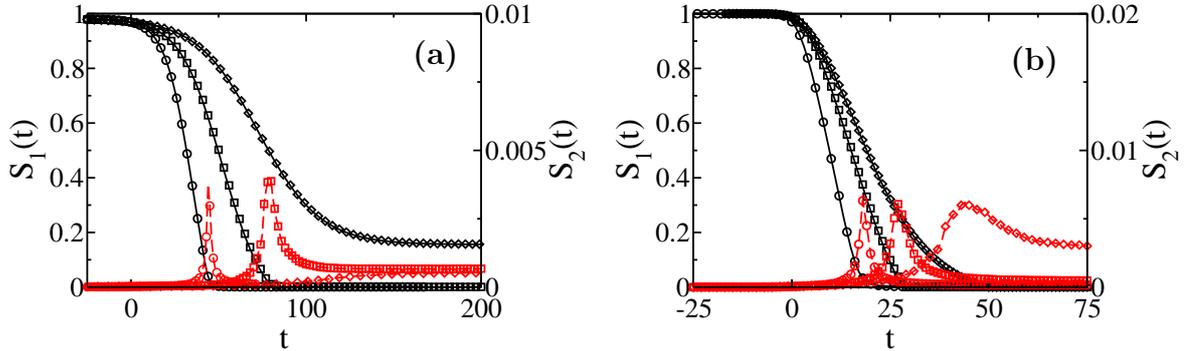

\centering
\vspace{1.0cm}
  \begin{overpic}[scale=0.24]{Fig03.eps}
    \put(70,50){{\bf{(a)}}}
  \end{overpic}\hspace{0.25cm}
  \begin{overpic}[scale=0.24]{Fig04.eps}
    \put(70,50){\bf{(b)}}
  \end{overpic}\vspace{0.25cm}
\caption{Time evolution of the size of the susceptible giant cluster
  $S_1(t)$ (black) and the second susceptible component $S_2(t)$
  (red), for $N=10^5$, $t_r=20$, $\beta=0.05$ ($T=0.64$) without
  intervention ($\bigcirc$) and with the ISD strategy for
  $\sigma=0.50$ and $t_b=1$ ($\square$, $T_{\sigma}=0.50$) and $t_b=2$
  ($\lozenge$, $T_{\sigma}=0.41$) with $\langle k \rangle=4$ in an ER
  network ($T^{*}=0.46$ and $T_c=0.25$) (a) and a SF with
  $\lambda=2.63$, $k_{min}=2$ and $k_{max}=500$ ($T^{*}=0.40$ and
  $T_c=0.05$) (b). The symbols correspond to the simulations and the
  solid lines are obtained from Eq.~(\ref{EqSusTotal2}). The dashed
  lines of the $S_{2}(t)$ curves are used as guides for the
  eyes. (Color online).}\label{Fig_Color_S1}
\end{figure}

In Fig.~\ref{Fig_Color_S1} we plot the time evolution of the size
$S_{1}(t)$ of the GSC, for ER and SF networks with the same average
degree, obtained from the theory and the simulations, where we call
``size'' to the fraction of nodes belonging to a cluster or component.
Remarkably, from the plots we can see the total agreement between the
theory and the simulations which shows that a complete description of
the time evolution of the size of the GSC can be given in this
adaptive strategy using theoretical tools~\cite{Exxon_Phys}.  On the
other hand, we also plot the size of the second biggest susceptible
cluster $S_{2}(t)$, in order to show the time at which the phase
transition develops on the GSC. This critical time is an important
quantity since it suggests that an intensification of the strategy is
needed in order to prevent or delay the destruction of the functional
GSC.

\begin{figure}[H]
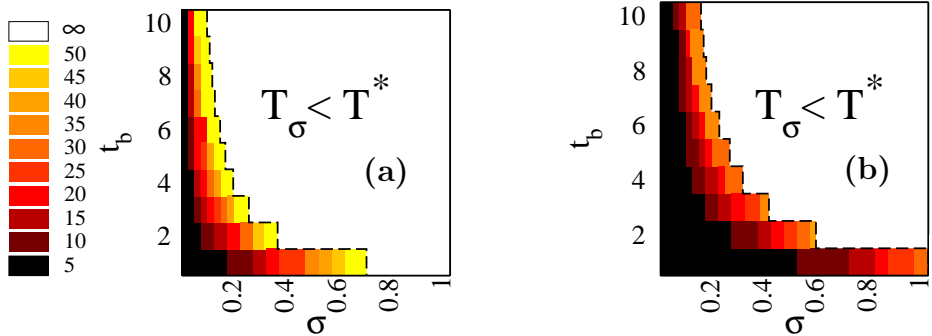

\centering
\vspace{1.0cm}
  \begin{overpic}[scale=0.24]{Fig05.eps}
    \put(80,35){{\bf{(a)}}}
  \end{overpic}\hspace{1.4cm}
  \begin{overpic}[scale=0.24]{Fig06.eps}
    \put(70,40){\bf{(b)}}
  \end{overpic}\vspace{0.25cm}
\caption{Phase diagram $t_b-\sigma$ displaying the increasing $\Delta
  t_c$ of the critical time from the case without intervention, for
  different values of $\sigma$ and $t_b$, $N=10^5$, $t_r=20$,
  $\beta=0.05$ with $\langle k \rangle=4$ in an ER network
  ($T^{*}=0.46$) (a) and a SF with $\lambda=2.63$, $k_{min}=2$ and
  $k_{max}=500$ ($T^{*}=0.40$) (b). The lowest increasing is in black
  color, and the highest in white color. The dashed lines correspond
  to the case $T_{\sigma}=T^{*}$ (Color online).}\label{Fig_Cuadros}
\end{figure}

As shown in the figure, the strategy increases the critical time $t_c$
compared to the case without mitigation strategy ($\sigma=0$), due to
the fact that as $t_{b}$ and $\sigma$ increases, the effective
transmissibility $T_{\sigma}$ decreases together with the number of
links traversed by the disease, protecting the GSC during a greater
period of time. In order to quantify $t_{c}$, we measure the position
of the peak of $S_{2}(t)$ (see Fig.~\ref{Fig_Color_S1}), which
corresponds to the average time at which the GSC is destroyed. In
Fig.~\ref{Fig_Cuadros} we plot in the plane $t_b-\sigma$, $\Delta
t_{c}=t_{c}(\sigma)-t_{c}(\sigma=0)$ for different values of $\sigma$
and $t_{b}$, obtained from the simulations.

From Fig.~\ref{Fig_Cuadros} we observe that the finite values of
$\Delta t_{c}$ only correspond to $T_{\sigma}>T^{*}$ [see
  Eq.~(\ref{TestrellaFinal})]. We can also see that the value of
$\Delta t_{c}$ increases when $T_{\sigma}\to T^{*}$ from above,
because below $T^{*}$ the GSC is never destroyed. In SF networks, this
increasing is less sensitive to the ISD strategy than in ER networks,
which is expected since in the former, the infected nodes with high
connectivities tend to enhance the disease spreading, decreasing
$S_{1}(t)$ at earlier times. However, there is a large range of values
of $t_b$ and $\sigma$ that allow to protect the giant susceptible
cluster for diseases with not too high values of $\beta$~\cite{Kit_01,
  Val_01}. The ability to decrease the transmissibility of the disease
spreading is an advantage of the ISD strategy for the susceptible
network since it gives to the public health authorities more time to
implement other policies, such as vaccination and to alert the health
services, before the functional network composed by healthy people
disappears.

\section{Delayed implementation of the ISD strategy}\label{SecDel_Straq}

While most of the adaptive strategies are applied at the beginning of
the disease spreading, in real situations, a mitigation strategy is
rarely implemented when the first infected
appears~\cite{Ban_01,Sat_01} because at this stage the health
authorities ignore the severity of the disease or they are cautious to
declared the alert of an epidemic since any strategy could affect the
functional susceptible network. In this section, we study the
performance of the strategy when it is implemented at a different time
after the beginning of the disease spreading.

In order to obtain the evolution equations of the delayed ISD
strategy, we consider that this strategy is implemented when a
fraction $x$ of the population is not susceptible.

If we denote $J(t)=I(t)+R(t)$, where $J(t)$ is the fraction of the
population that was reached by the spread (incidence curve), when
$J(t)<x$, the evolution of the disease spreading on top of a static
network is governed by the Eqs.~(\ref{delta_theta}),
(\ref{deltaPhiS}),~(\ref{EqInfec})-(\ref{EqSusTotal1}), except for
$\Phi_{I}(t)$, that is given by~\cite{Val_02},
\begin{eqnarray}
\Delta{\Phi}_{I}(t)&=&-\beta\Phi_{I}(t)-\Delta{\Phi}_{S}(t)+(1-T)\Delta{\Phi}_{S}(t-t_r)\label{deltaPhiApp}.
\end{eqnarray}
When $J(t)\geq x$ at $t=\tau$, the ISD strategy is implemented and the
evolution of the spreading follows
Eqs.~(\ref{delta_theta})-(\ref{deltaPhi}),~(\ref{EqInfec})-(\ref{EqSusTotal1}). However,
note that when $J(\tau)=x$, there is a change in the transmissibility,
since just after the disease reaches $x$ percent of the population,
there are still some active links that appeared before $t=\tau$, which
we call ``old links''. As a consequence, the effective probability
that these links would be traversed at early stages is not $T$ neither
$T_{\sigma}$. In order to incorporate this issue in the evolution
equations, first we have to compute $\Phi_{I, z}\equiv
(1-\beta)^{t_r-z}\Delta\Phi_{S}(\tau-t_r+z)$, \ie, the amount of old
active links which appeared $t_r-z$ times units before $t=\tau$. This
is equivalent to have a smaller recovery time $z$ (with $1\leq z\leq
t_r-1$) instead of $t_r$, in the evolution equations of the ISD
strategy \footnote{Note that $z$ must be less than $t_r$ because an
  old link spend at least one time unit in the period without
  strategy.}. As a consequence, the probability that these old links
have not yet transmitted the disease at time $t=\tau+j$ with $j\leq z$
is $\Omega_j$ (see Eq.~(\ref{Ec.omesuper})). Here, we consider
$\Omega_j=0$ for $j<1$ and $j>t_r-1$ for the case of old links. Then,
the equation of $\Phi_{I}(t)$ for the delayed ISD strategy needs to
take into account the old active links, thus
\begin{eqnarray}
\Delta{\Phi}_{I}(t)&=&-\beta\Phi_{I}(t)-\Delta{\Phi}_{S}+(1-\beta)(1-\sigma)\Omega_{t_{r}}\Delta{\Phi}_{S}(t-t_{r})\Theta_{1}+ \nonumber \\
&&+(1-\beta)(1-\sigma)\Omega_{t-\tau+1}\Phi_{I, t-\tau+1}(1-\Theta_1)+ \Psi^{*}(t),\label{deltaPhiSApp11}
\end{eqnarray}
with
\begin{eqnarray}\label{reconection_termApp}
\Psi^{*}(t)&\equiv&-\sigma(1-\beta)\Phi_{I}(t)+\sigma(1-\beta)\Phi_{I}(t-t_{b})\Theta_0\nonumber\\
&&+\left(\sum_{i=0}^{t_{b}}\sigma(1-\beta)\Omega_{t_r-t_b+i}\Delta{\Phi}_{S}(t-i-t_{r})\right)\Theta_1\nonumber\\
&&+\left(\sum_{z=t-\tau-t_b}^{t-\tau}\sigma(1-\beta)\Omega_{t-\tau-t_b}\Phi_{I,z}\right)(1-\Theta_2),
\end{eqnarray}
where $\Theta_0$, $\Theta_1$ and $\Theta_2$ are the Heaviside functions given by
$$
\Theta_0 =  \left\{
\begin{array}{cl}
0\;\;\; &\mbox{if }\;\;\; t<\tau+t_b-1 \\
1\;\;\; &\mbox{otherwise }
\end{array}\right.\\,
\Theta_1 =  \left\{
\begin{array}{cl}
0\;\;\; &\mbox{if }\;\;\; t<\tau+t_r-1 \\
1\;\;\; &\mbox{otherwise }
\end{array}\right.\\,
$$

$$
\Theta_2 =  \left\{
\begin{array}{cl}
0\;\;\; &\mbox{if }\;\;\; t<\tau+t_r+t_b-1 \\
1\;\;\; &\mbox{otherwise }
\end{array}\right.
$$

The first two terms of Eq.~(\ref{deltaPhiSApp11}) are the same as
those in Eq.~(\ref{deltaPhi}). The third term corresponds to the
recovery of the active links which appeared at time $t\geq
\tau-1$. This term is different from zero only after $t_r-1$ time
units since $\tau$, and has the same form as the third term of
Eq.~(\ref{deltaPhi}). The fourth term describes the recovery of the
infected individuals of old active links. Note that this term, has the
same form as the previous one, in which $\Phi_{I,t-\tau+1}$ plays
the role of $\Delta{\Phi}_{S}(t-t_{r})$. Finally the fifth term
corresponds to the breaking-reconnection term that is similar to
Eq.~(\ref{reconection_term}).

\begin{figure}[H]
\centering
\vspace{0.3cm}
 \includegraphics[scale=0.6]{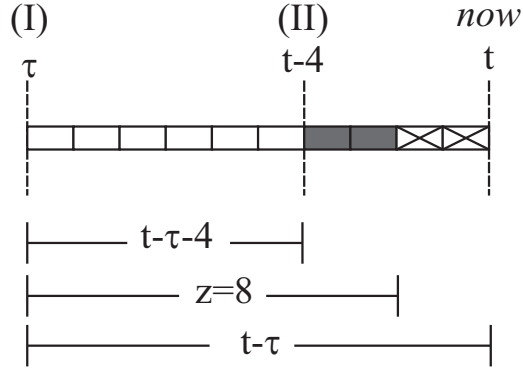}
  \vspace{0.5cm}
\caption{Schematic of the correction of the restoring term of
  Eq.~(\ref{reconection_termApp}) at time $t$, for an old active link
  with $z=8$ and $t_b=4$ that breaks two units time before the
  infected node recovers. Each cell corresponds to a time unit.  The
  white cells represent the time unit at which an active link between
  the susceptible and the infected node exists, the gray ones
  denote the disconnection period and the crosses correspond to
  the case where the infected is recovered. The moment $\tau$ at which
  the ISD strategy is implemented, is represented by (I), and the
  moment when this active link, at time $t-t_b$, breaks for the next $t_b$
  time units is denoted by (II). The fraction of active links
  represented in the plot is proportional to
  $(1-\beta)\sigma\Omega_{t-\tau-4}$.}\label{Schem_delay}
\end{figure}

In Eq.~(\ref{reconection_termApp}), the first two terms are similar
as those in Eq.~(\ref{reconection_term}). The third term corresponds
to the correction on the reconnection term due to the recovery of
infected individuals of the active links that appear after
$t=\tau$. On the other hand, the last term corresponds to the
correction due to the recovery of the infected individuals of the old
links. In Fig.~\ref{Schem_delay} we show a schematic of this term.

\begin{figure}[H]
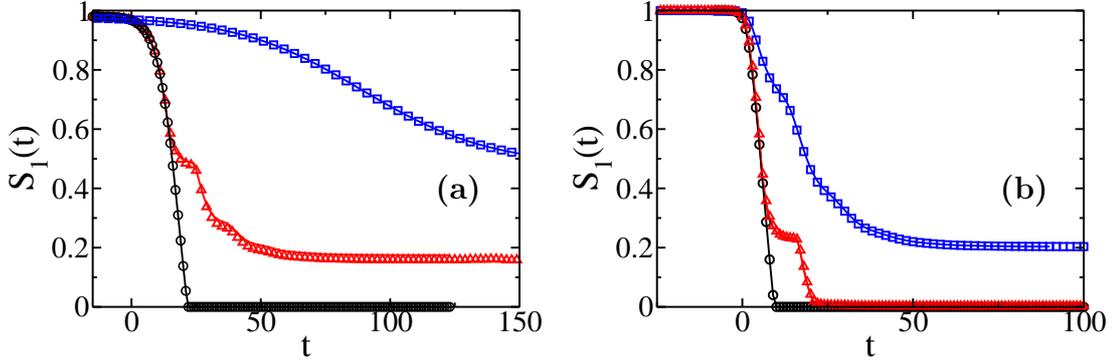

\centering
\vspace{1.0cm}
  \begin{overpic}[scale=0.26]{Fig08.eps}
    \put(80,30){{\bf{(a)}}}
  \end{overpic}\hspace{0.25cm}
  \begin{overpic}[scale=0.26]{Fig09.eps}
    \put(80,30){\bf{(b)}}
  \end{overpic}\vspace{0.25cm}
\caption{$S_1(t)$ for a disease with $t_r=20$, $\beta=0.10$ ($T=0.88$)
  and $N=10^5$ when the ISD strategy (with $\sigma=0.50$ and $t_b=10$,
  \ie, $T_{\sigma}=0.33$) is applied at different moments: never
  (black, $\bigcirc$), after the disease reaches 25\% ($x=0.25$) of
  the population (red, $\triangle$), and from the beginning (blue,
  $\square$) for an ER network with $\langle k \rangle =4$ (a) and a
  SF network with $\lambda=2.63$, $k_{min}=2$ and $\langle k \rangle
  =4$ (b). The symbols correspond to the simulations with 100 network
  realizations, and the solid lines correspond to the theoretical
  solutions. In ER network, for the case with strategy applied from
  the beginning $S_{1}(t\to \infty)= 0.47$. (Color
  online).}\label{Distintos_momentos}
\end{figure}

In Fig.~\ref{Distintos_momentos} we plot $S_1(t)$ as a function of $t$
for ER and SF networks, when the ISD strategy is applied after the
disease reaches a fraction $x$ of the population.

As shown in the figure, the strategy can protect a GSC, even when it
is applied after a large fraction of the population is infected. Note
that the delayed strategy does not change the behavior of the order
parameter $S_{1}(t)$ with $\Phi_S(t)$, thus, the critical value of
$\Phi_{Sc}$ at which the giant susceptible cluster is destroyed
remains invariant (see Appendix). On the other hand, the strategy
protects a smaller fraction of individuals in the functional network in
SF than in an ER networks, because in the former the dilution
process of the susceptible network is more efficient due to the nodes
with high connectivities (see Sec.~\ref{Node_Void}). However, a delayed
intermittent connection strategy can still protecting the GSC by
increasing $t_c$, and thus mitigating the demand of health services
and protecting the functional network for an extended period of time.

\section{Conclusions}\label{SecConc}
In this paper, we study how individuals based on local information,
contribute to halt the epidemic spreading implementing an ISD strategy
during a disease spreading. This model where the healthy individuals
avoid contact with the infected ones intermittently, allows to mimic a
behavioral response of individuals who try to protect themselves from
the disease but also try to preserve their closer contacts, such as
friendship and working partners.

Using an edge-based compartmental model combined with percolation
theory, we show that this strategy increases the critical time at
which the susceptible giant component is destroyed, giving more time
to the health authorities to implement other policies against the
disease spreading. We also study a more realistic scheme in
which the strategy is delayed, and found that it also protects the
functional network. We show that the dilution of the GSC for the
delayed strategy can also be described by percolation theory. Our
theoretical framework are fully supported by extensive simulations.

We believe that to focus on the susceptible network and its
topological properties, instead that on the infected network, provides
a novel description that could be useful for the health services to
develop new strategies to protect the society and the economy of a
region. Finally, this complementary view of an epidemic spreading
could give a new criterion to evaluate the effectiveness of any
strategy.

\section*{Acknowledgements}
We thank UNMdP and FONCyT (Pict 0293/2008) for financial support.

\appendix 

\section{Node Void Percolation in a delayed ISD strategy}

A feature of the delayed strategy is that the disease spreads with two
different values of the transmissibility ($T$ before applying the
strategy and $T_{\sigma}$ during the strategy), and thus the final
epidemic size cannot be related with a unique value of the control
parameter $T$.

\begin{figure}[H]
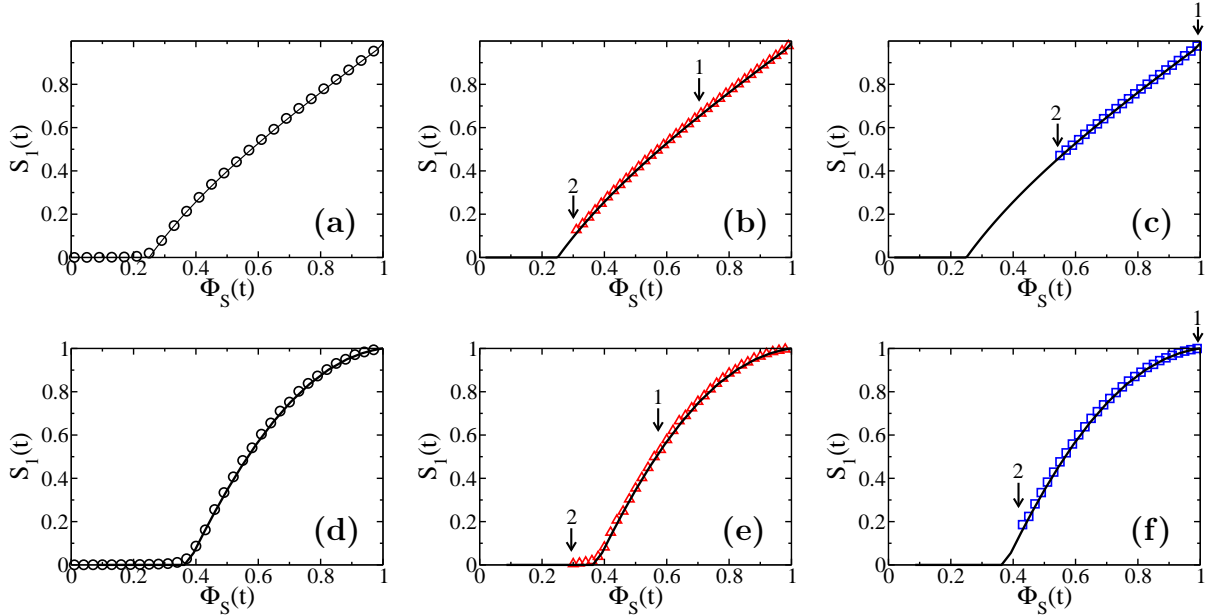

\centering
\vspace{1.0cm}
  \begin{overpic}[scale=0.19]{Fig10.eps}
    \put(80,20){{\bf{(a)}}}
  \end{overpic}\hspace{0.25cm}
  \begin{overpic}[scale=0.19]{Fig11.eps}
    \put(80,20){\bf{(b)}}
  \end{overpic}\hspace{0.25cm}
  \begin{overpic}[scale=0.19]{Fig12.eps}
    \put(80,20){{\bf{(c)}}}
  \end{overpic}\hspace{0.25cm}
  \begin{overpic}[scale=0.19]{Fig13.eps}
    \put(80,20){\bf{(d)}}
  \end{overpic}\hspace{0.25cm}
  \begin{overpic}[scale=0.19]{Fig14.eps}
    \put(80,20){{\bf{(e)}}}
  \end{overpic}\hspace{0.25cm}
  \begin{overpic}[scale=0.19]{Fig15.eps}
    \put(80,20){\bf{(f)}}
  \end{overpic}\vspace{0.25cm}
\caption{$S_1(t)$ as a function of $\Phi_S(t)$ for a disease with
  $t_r=20$, $\beta=0.10$ ($T=0.88$) and $N=10^5$ (with $\sigma=0.50$
  and $t_b=10$, \ie, $T_{\sigma}=0.33$) when the ISD strategy is
  applied at different moments: never (black, $\bigcirc$), after the
  disease reaches 25\% ($x=0.25$) of the population (red,
  $\triangle$), and from the beginning (blue, $\square$) for an ER
  network with $\langle k \rangle =4$ [Figs. a), b), c)] and a SF
  network with $\lambda=2.63$, $k_{min}=2$ and $\langle k \rangle =4$
  [Figs. d), e), f)]. The symbols correspond to the simulations with
  100 network realizations and the solid black lines are the solution
  of Eqs.~(\ref{delta_theta})-(\ref{EqSusTotal1}) for the case without
  strategy. The time $t$ is implicit, and increases when $\Phi_{S}(t)$
  decreases. The arrows with label $1$ represent the moment at which
  the strategy is applied, and the ones with label $2$
  correspond to the steady state for the ISD strategy. (Color
  online).}\label{Distintos_momentosPhi}
\end{figure}

In order to study the effect of the delayed strategy on the dilution
of the susceptible network, we analyze the relation between $S_1(t)$
and $\Phi_{S}(t)$ during the disease spreading. In
Fig.~\ref{Distintos_momentosPhi} we plot $S_1(t)$ as a function of
$\Phi_{S}(t)$, obtained from the theory (see
Eqs.~(\ref{delta_theta})-(\ref{EqSusTotal1})) and the simulations,
when the ISD strategy is applied after the disease
reaches a fraction $x$ of the population, for ER and SF networks.

Even thought there is not a fixed transmissibility, as shown in
Fig.~\ref{Distintos_momentosPhi}, the delayed strategy does not change
the behavior of the order parameter $S_{1}(t)$ with $\Phi_S(t)$ found
in the case without intervention~\cite{Val_02}. This result is
expected since although the speed at which the susceptible giant
cluster is diluted changes, the size of the GSC at time $t$ depends on
the amount of susceptible nodes removed by node void percolation, and
not directly on the transmissibility. As a consequence, the relation
between $S_{1}(t)$ and $\Phi_S(t)$ holds even for a varying
transmissibility. Then, the critical value of $\Phi_{Sc}$ at which the
GSC is destroyed remains invariant. This implies that in a realistic
scenario in which the transmissibility is varying, if $\Phi_{S}(t)$ is
approaching to $\Phi_{Sc}$, the GSC is near to being destroyed. Thus
the value of the distance to the criticallity of the susceptible
network $\Phi_{S}(t)-\Phi_{Sc}$, could be a crucial information for
the authorities to decide if more aggressive health policies are
needed to halt the epidemic spreading and to protect the functional
network.

 \bibliography{bibpaper_corr}

\begin{thebibliography}{10}
\bibitem{Exxon_06}
Note that $\Phi_{S}=1-f_{\infty}$, where $f_{\infty}$ is the probability that
  the disease leads to a giant infected component following a link, with
  $f_{\infty}=1-G_{1}[1-T\;f_{\infty}]$, thus $\Phi_{S}=G_{1}[1-T(1-\Phi_S)]$.
\bibitem{Exxon_Phys}
Also we obtain a total agreement between the theory and the simulations above
  the epidemic threshold for $t_r\geq 2$.
\end{thebibliography}
\end{document}